\documentclass[twocolumn,fleqn,usenatbib,useAMS,epjc3]{svjour3}
\usepackage{amsmath,amssymb,tikz,pgfplots}

\begin{document}

\title{The cosmological background and the ``external field'' in Modified Gravity (MOG)}

\author{J. W. Moffat$^{\star}$ and
V. T. Toth$^\star$\\~\\
{\rm
\footnotesize
$^\star$Perimeter Institute for Theoretical Physics, Waterloo, Ontario N2L 2Y5, Canada}}

\maketitle

\begin{abstract}
We investigate the contributions of the Friedmann--Lema\^itre--Robertson--Walker metric of the standard cosmology as an asymptotic boundary condition on the first-order approximation of the gravitational field in Moffat's theory of modified gravity (MOG). We also consider contributions due to the fact that the MOG theory does not satisfy the shell theorem or Birkhoff's theorem, resulting in what is known as the ``external field effect'' (EFE). We show that while both these effects add small contributions to the radial acceleration law, the result is orders of magnitude smaller than the radial acceleration in spiral galaxies.
\end{abstract}

\section{Introduction}

The theory of Scalar-Tensor-Vector Gravity (STVG; \cite{Moffat2006a}), also known by the acronym MOG (for MOdified Gravity), is a modified theory of gravitation based on an action principle that incorporates, in addition to the metric of general relativity, a massive vector field that couples to matter universally. The theory has been used successfully to model a range of phenomena that are usually attributed to hypothetical collisionless dark matter, including galaxy rotation curves, the matter power spectrum and the acoustic power spectrum of the cosmic microwave background.

Recently in the literature, in the context of other modified theories of gravitation, the so-called ``external field effect'' (EFE; \cite{BekensteinMilgrom1984,McGaugh2020}) has been discussed. This prompted us to investigate whether or not such an effect is present in MOG and if so, the magnitude of its contribution to galaxy dynamics.

We specifically investigate two possible sources of such an effect: contributions due to the non-flat boundary conditions present in a Friedmann--Lema\^itre--Robertson--Walker (FLRW) cosmology, and contributions due to violations of Birkhoff's theorem \cite{BIRKHOFF1923,Weinberg1972} in the MOG theory.

To investigate the effects of the cosmological background, we adopt McVittie's metric \cite{McVittie1933}, which can be used to analyze the contributions of the FLRW background to the Schwarzschild solution. Regarding the EFE, we rely on earlier results \cite{MOG2021a} as we demonstrate that in MOG, although the effect arises naturally as a consequence of the theory not obeying Birkhoff's theorem, its magnitude is too small to have a significant impact on, e.g., the predicted behavior of galaxies and stellar clusters of various sizes. Of course the MOG theory neither requires nor relies on an EFE to model galaxy rotation curves, cluster dynamics and cosmology. The existence of an EFE in the theory can still be of interest, as (among other things) it provides an example of how a properly formulated relativistic theory of gravitation can yield such an effect without relying on unjustified, {\em ad hoc} assumptions.

\section{The MOG gravitational potential and acceleration}

Derived from the field equations and Lagrangian of the MOG theory, the MOG modified gravitational potential for a point source of mass $M$, in the case of weak gravitational fields and low velocities is given by \cite{Moffat2006a}:
\begin{align}
\Phi(r)=-\frac{G_NM}{r}[1+\alpha(1-e^{-\mu r})],
\end{align}
where $G_N$ is Newton's constant for gravitation and $r$ is the distance from the gravitating mass. The corresponding radial acceleration law is obtained as usual, as the gradient of the potential:
\begin{equation}
\label{acceleration}
a(r)=-\frac{G_NM}{r^2}[1+\alpha-\alpha e^{-\mu r}(1+\mu r)].
\end{equation}
The parameters $\alpha$ and $\mu$ can be estimated by introducing constraints to the theory \cite{Moffat2007e}, or by fitting them to the rotation or velocity dispersion profiles of galaxies or clusters of galaxies \cite{MoffatRahvar2013,Israel2016,GreenMoffat2019,DavariRahvar2020,Moffat2020a}. Typical values, for example, are $\alpha\sim 10$, $\mu\sim (50~{\rm kpc})^{-1}$.

\section{The McVittie metric}

McVittie's metric \cite{McVittie1933,SKMHH2003} describes a compact, spherically symmetric source of gravitation, with mass $M$ embedded in a Friedmann--Lema\^itre--Robertson--Walker\\ (FLRW) cosmological background. In comoving coordinates, McVittie's metric can be written as
\begin{align}
ds^2 = & \frac{\left[1 - \dfrac{GM}{2c^2aR}\right]^2}{\left[1 + \dfrac{GM}{2c^2aR}\right]^2}~c^2dt^2 \nonumber\\
&{}- a^2\left[1 + \frac{GM}{2c^2aR}\right]^4 (dR^2 + R^2 d\Omega^2),
\end{align}
where $a=a(t)$ is the dimensionless scale factor of the FLRW universe and $G$ is the gravitational constant.

Rewriting the metric using $r=aR$ and $\Phi=-GM/r$, in the limit $\Phi\ll c^2$ and (using $H=\dot{a}/a$ as the Hubble-parameter) $Hr\ll c$ (both of which are applicable when considering galaxy dynamics), we obtain the following approximate form:
\begin{align}
ds^2 = & \left(1 + 2\frac{\Phi}{c^2}-\frac{H^2r^2}{c^2}\right)c^2dt^2+2\frac{Hr}{c}dr~ c~dt\nonumber\\
&{}- \left(1 - 2\frac{\Phi}{c^2}\right) d{\bf r}^2.
\end{align}

In the large $aR$ limit, this metric approaches the Schwarzschild solution. As such, it also matches the spherically symmetric, static MOG vacuum solution beyond the range of the vector field, $aR\gg\mu^{-1}$.

As we are well within the regime of the linear approximation, we expect this form to remain valid at the same level of approximation when $\Phi$ represents the MOG gravitational potential.

It is, of course, true that within the range of the vector force, $r\lesssim \mu^{-1}$, the MOG field equations are quite different from the field equations of general relativity. However, at cosmological distances, $r\gg\mu^{-1}$, the contribution of the vector force vanishes and the theory becomes indistinguishable from Einstein's theory, apart from the difference in the value in the gravitational constant. The vector force, in effect, governs the internal dynamics of the gravitating object but does not affect the gravitational field over very large distances. Conversely, while the internal dynamics of a MOG galaxy are determined by the full set of field equations, in the nonrelativistic approximation they reduce to the MOG acceleration law. The (small) perturbation of this acceleration due to cosmological matter can be estimated using the McVittie metric, since in the linear approximation, the magnitude of the perturbation ($\propto H^2r^2/c^2$) is not affected by the local matter density or local fields.

Also note that while McVittie's metric becomes singular at its horizon \cite{Ferraris1996}, the horizon scale is several orders of magnitude smaller than our region of interest, the scale of a typical galaxy. At a much shorter range, we would expect the MOG field equations to dominate, with any cosmological contribution becoming negligible, yielding the Kerr--Newman type MOG solutions described elsewhere (see, e.g., \cite{Moffat2015a}).

The $dr~dt$ mixed term represents a gravitomagnetic contribution. Although its magnitude is not insignificant, as it is a purely radial term, it has vanishing curl and thus it has no impact on the dynamics of the host galaxy.

In contrast, the contribution to the $dt^2$ term, characterized by $H^2r^2/c^2$, though small, represents a radial potential that can alter the dynamics of a galaxy. Notably, the magnitude of this term is independent of the mass of the galaxy. The corresponding radial acceleration is given by
\begin{align}
a_H(r)=2H^2r=1.82\times 10^{-14}h_{75}^2\left(\frac{r}{50~{\rm kpc}}\right)~\frac{\rm m}{{\rm s}^2},\label{eq:aH}
\end{align}
where $h_{75}=H/(75~{\rm km}/{\rm s}/{\rm Mpc})$ is defined as usual.

This acceleration is four orders of magnitude smaller than the centrifugal acceleration of our solar system in the Milky Way, which can be written as
\begin{align}
\hskip -1em
a_c(r)=\frac{v_c^2}{r}=1.96\times 10^{-10}\left(\frac{v}{220~{\rm km}/{\rm s}}\right)^2\left(\frac{8~{\rm kpc}}{r}\right)\frac{\rm m}{{\rm s}^2},
\end{align}
using nominal values that approximately characterize the orbital motion of our own solar system in the Milky Way.

\section{MOG Acceleration}

Applying the MOG acceleration law (\ref{acceleration}) to an extended, spherically symmetric mass distribution, we arrive at the following integral form \cite{MOG2019a}\footnote{In Eq.~(14) in \cite{MOG2019a}, the symbol $G$ should have been $G_N$.}:
\begin{align}
\label{eq:a_r}
a(r)=&
-\int_0^{r}dr'\dfrac{2\pi G_Nr'}{\mu r^2}\rho(r')\\
&\hskip -0.25in{}\times\bigg\{2(1+\alpha)\mu r'+\alpha(1+\mu r)\left[e^{-\mu(r+r')}-e^{-\mu(r-r')}\right]\bigg\}\nonumber\\
&{}-\int_{r}^\infty dr'\dfrac{2\pi G_Nr'}{\mu r^2}\rho(r')\nonumber\\
&{}\times\alpha\bigg\{\big[1+\mu r\big]e^{-\mu(r'+r)}-\big[1-\mu r\big]e^{-\mu(r'-r)}\bigg\}.\nonumber
\end{align}

This form reveals a peculiar property of the MOG theory: as the acceleration law is not an inverse-$r^2$ law, the theory does not satisfy the shell theorem or Birkhoff's theorem \cite{BIRKHOFF1923,Weinberg1972}. The second integral in the sum above is nonzero, and represents the contribution of the gravitational acceleration on a test particle at distance $r$ from the center of the spherical distribution, by shells of matter at radius $r$ and above.

In light of recent discussions \cite{McGaugh2020} on the importance of the ``external field effect'' in certain modified gravity formulations, we found it instructive to estimate the magnitude of this contribution, to see what, if any, effect it has on the equations of motion in typical astrophysical scenarios.

Evaluating the second integral in (\ref{eq:a_r}) would make sense in Minkowski spacetime, but not in the spacetime of an FLRW universe. The gravitational influence of distant matter must take into account that we see distant matter at a time when the average density of the universe was much higher.

Such considerations are obviously model-dependent, especially when it comes to the very early universe. Instead of elaborating on such model dependencies, in order to obtain a reasonable order-of-magnitude estimate of the contribution of this term, we just evaluate the integral up to the surface of last scattering, and assume a pressureless equation of state ($\rho\propto a^{-3}$ where $a$ is the scale factor of the expanding universe) and a uniform rate of expansion. In short, we evaluate

\begin{align}
a_{\rm ext}(r)=&\int_{r}^{r_1} dr'\dfrac{2\pi G_Nr'}{\mu r^2}\rho(r')\label{eq:aext}\\
&{}\times\alpha\bigg\{\big[1+\mu r\big]e^{-\mu(r'+r)}-\big[1-\mu r\big]e^{-\mu(r'-r)}\bigg\}\nonumber
\end{align}
where we now use $r_1$ to denote the light travel distance to the surface of last scattering. Denoting the light travel time radius of the surface of last scattering with $r_2$, we estimate the density using
\begin{align}
\rho(r') = \frac{(r_1+r_2)^3}{(r_1+r_2-r')^3}\rho_0,
\end{align}
where $\rho_0=3H_0^2/\big(8\pi (1+\alpha)G_N\big)$ is the MOG critical density at the present epoch and $G=(1+\alpha)G_N$. The expression in the numerator is proportional to the scale factor at the present epoch, $a\propto r_1+r_2$, whereas the term in the denominator is approximately proportional to the scale factor at the time that corresponds to $r'$, $a(r')\propto r_1+r_2-r'$.

Using $r_1=13.8\times 10^9$~ly, $r_2=385,000$~ly and $H_0=70~{\rm km}/{\rm s}/{\rm Mpc}$, we obtain, at a typical galactocentric distance of $r=50$~kpc and using $\mu^{-1}=50$~kpc, $\alpha=10$,
\begin{align}
a_{\rm ext}(50~{\rm kpc})\sim 2.9\times 10^{-15}~{\rm m}/{\rm s}^2.
\end{align}
Evaluating this expression numerically, we find that the result is only weakly sensitive to specific choices of $r_1$, $r_2$ and $r$ and the order of magnitude of the result remains the same. Therefore, though an approximation, it is a reliable estimator for the magnitude of the nonrelativistic contribution to the galactocentric acceleration by distant matter due to the Birkhoff-violating nature of the MOG theory. We are, of course, ignoring matter behind the surface of last scattering, as its contribution is difficult to calculate and the nonrelativistic approximations used here are not applicable in this regime, but it can be reasonably assumed not to change this result substantially. In particular, we find that the galactocentric acceleration due to distant matter is several orders of magnitude less than the effect required to substantially influence the rotation of even large galaxies.

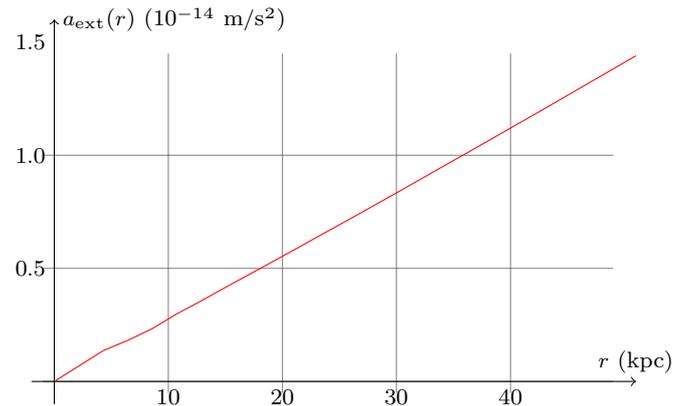
\begin{figure}

\begin{center}
\begin{tikzpicture}[domain=0.01:5.1, scale=1.5]
%\draw[very thin,color=gray] (-0.1,-0.1) grid (4.9,3.9);
\draw[very thin,color=gray] (-0.1,-0.1) grid (4.9,2.9);
\draw[->] (-0.2,0) -- (5.1,0) node[above] {$r~({\rm kpc})$};
\draw[->] (0,-0.2) -- (0,3.2) node[right] {$a_{\rm ext}(r)~(10^{-14}~{\rm m}/{\rm s}^2)$};
\draw (0,1) node[left] {$0.5$};
\draw (0,2) node[left] {$1.0$};
\draw (0,3) node[left] {$1.5$};
\draw (1,0) node[below] {$10$};
\draw (2,0) node[below] {$20$};
\draw (3,0) node[below] {$30$};
\draw (4,0) node[below] {$40$};
\draw[color=red] plot (\x,{2*(0.3171*\x-0.01296*(1+0.2*\x)*(exp(-0.4*\x)*0.2*\x+0.2*\x+exp(-0.4*\x)-1)*625/(\x*\x))});
\end{tikzpicture}
\end{center}
\caption{\label{fig:a}Radial acceleration in a spherically symmetric configuration due to the FLRW cosmological background in the MOG theory. Deviations from Birkhoff's theorem in the theory result in a barely observable perturbation due to matter outside of radius $r$. Nominal values of the parameters in Eqs.~(\ref{eq:aH}) and (\ref{eq:aext}) were used.}
\end{figure}

\section{Conclusions}

As it has been known since at least 1933, the Schwarzschild solution for the gravitational field of a compact gravitating object must be modified when its boundary conditions are those of an expanding FLRW spacetime as opposed to the flat Minkowski metric. In the present paper we estimated the magnitude of the resulting radial acceleration and ascertained that it is orders of magnitude less than typical values for the centrifugal acceleration of objects in a large spiral galaxy, such as our Milky Way.

We also estimated acceleration due to shells of matter surrounding a galaxy. Because MOG predicts an acceleration law that does not follow the inverse square relationship, the theory violates Birkhoff's theorem. Acceleration in a spherically symmetric distribution of matter is dependent not just on matter inside but also matter outside the shell on which a test particle resides, leading to a possible External Field Effect. The magnitude of this acceleration is even smaller than the acceleration due to spacetime expansion.

Consequently, we established that neither the background cosmology nor violations of Birkhoff's theorem alter the dynamics of rotating galaxies by any observable amount in the MOG theory. Our results also apply as a matter of course to the dynamics of galaxies under standard general relativity, with or without dark matter.

\begin{acknowledgement}

This research was supported in part by Perimeter Institute for Theoretical Physics. Research at Perimeter Institute is supported by the Government of Canada through the Department of Innovation, Science and Economic Development Canada and by the Province of Ontario through the Ministry of Research, Innovation and Science.
VTT acknowledges the generous support of Plamen Vasilev and other Patreon patrons.

\end{acknowledgement}

\bibliography{refs}
\bibliographystyle{spphys}

\end{document}